\documentclass{PoS}
\usepackage{epsfig}

\newcommand{\<}{\langle}
\renewcommand{\>}{\rangle}

\newcommand{\beq}{\begin{equation}}
\newcommand{\eeq}{\end{equation}}
\newcommand{\beqn}{\begin{eqnarray}}
\newcommand{\eeqn}{\end{eqnarray}}

\PoS{PoS(LAT2005)285}

\title{Cutoff effects in maximally twisted LQCD}

\ShortTitle{Cutoff effects in Mtm-LQCD}

\author{\speaker{Giancarlo Rossi}\\
        Dipartimento di Fisica, Universit\`a di  Roma ``{\it Tor Vergata}'' - Roma, Italy\\
        INFN, Sezione di Roma2 - Roma, Italy\\
        John von Neumann-Institut f\"ur Computing, NIC - Zeuthen, Germany\\
        E-mail: \email{rossig@roma2.infn.it}}

\author{Roberto Frezzotti{\thanks{From Sept. 1$^{st}$, 2005, 
        new affiliation and address: Universit\`a di Roma ``{\it Tor Vergata}'', 
        Dipartimento di Fisica,
        Via della Ricerca Scientifica -- 00133 Roma (Italy).
        $E$-$mail$: Roberto.Frezzotti@roma2.infn.it
        }}\\
        INFN, Sezione di Milano \\
        Dipartimento di Fisica, Universit\`a di Milano ``{\it Bicocca}'' - Milano, Italy\\
        \mbox{E-mail: \email{Roberto.Frezzotti@mib.infn.it}}}

\author{Guido Martinelli\\
        Dipartimento di Fisica, Universit\`a di Roma ``{\it La Sapienza}'' - Roma, Italy\\
        INFN, Sezione di Roma ``{\it La Sapienza}'' - Roma, Italy\\
        E-mail: \email{Guido.Martinelli@roma1.infn.it}}

\author{Mauro Papinutto\\
        John von Neumann-Institut f\"ur Computing, NIC - Zeuthen, Germany\\
        E-mail: \email{Mauro.Papinutto@desy.de}}

\abstract{The Symanzik analysis of correlators in lattice QCD with maximally 
twisted Wilson fermions reveals that there exist cutoff artifacts which tend to 
become large as the quark mass gets small. We show that these effects can be 
reduced to a negligible level by either introducing the clover term in the 
action or, as already suggested in the literature, by a suitable choice of the 
critical mass. Recent simulation data support these conclusions.}

\FullConference{XXIIIrd International Symposium on Lattice Field Theory\\
                  25-30 July 2005\\
                  Trinity College, Dublin, Ireland}

\begin{document}

\section{Introduction and main results}
\label{sec:INTRO}

Although in lattice QCD at maximal twist (Mtm-LQCD) O($a$) discretization 
effects (actually all O($a^{2k+1}$), $k\geq 0$ effects) are absent 
or easily eliminated~\cite{TM,FR1,FR2,FR4}, it turns out that correlators 
are affected by dangerous artifacts of relative order $a^{2k}$, $k\geq 1$, 
which are enhanced by inverse powers of the (squared) pion mass, as the latter 
becomes small. In fact, when analyzed in terms of the Symanzik expansion, 
lattice expectation values exhibit, as $m_\pi^2\to 0$, what we will call 
``infrared (IR) divergent'' cutoff effects with a behaviour of the form 
\beq
\<O\>\Big{|}^L_{m_q}=\<O\>\Big{|}^{\rm cont}_{m_q}
\Big{[}1+{\rm O}\Big{(}\frac{a^{2k}}{(m_\pi^2)^{h}}\Big{)}\Big{]}
\, ,\quad 2k\geq h\geq 1 \,\,(k,h \,\,{\rm integers})\, ,\label{ORD}\eeq 
where we have assumed that the lattice correlator 
admits a non-trivial continuum limit. Powers of $\Lambda_{\rm QCD}$ required 
to match physical dimensions are often understood in the following. 

We shall see that artifacts of the type~(\ref{ORD}) are reduced to terms 
that are at worst of order $a^{2}(a^2/m_\pi^2)^{k-1}$, $k\geq 1$, 
if the action is O($a$) improved {\it \`a la} Symanzik, or alternatively
the critical mass is chosen in some ``optimal'' way.

The idea that a suitable definition of critical mass exists which 
can lead to a smoothing out of chirally enhanced lattice artifacts or
perhaps be of help in getting improvement was already put forward in the 
context of chiral perturbation theory in refs.~\cite{SHWUNEW} 
and~\cite{AB}, respectively.

An important consequence of our analysis is that the strong 
(order of magnitude) inequality $m_q> a\Lambda^2_{\rm QCD}$,
invoked in ref.~\cite{FR1} 
can be relaxed to the weaker relation $m_q> a^2\Lambda^3_{\rm QCD}$,
before large cutoff effects are possibly 
met while lowering the quark mass at fixed $a$.
The works of refs.~\cite{SHWUNEW,AB}, and most recently
refs.~\cite{AB05,SH05}, which are based on lattice chiral 
perturbation theory, lead to essentially equivalent conclusions about 
cutoff effects in pion quantities in the parameter region 
$m_q> a^2 \Lambda^3_{\rm QCD}$. They also yield interesting 
predictions on the possible Wilson fermions phase scenarios~\cite{PS,SCO} 
and results, when $m_q$ is of order $a^2$ or smaller. 

A thorough discussion on the effectiveness of Mtm-LQCD in killing O($a$) 
discretization errors and the ability of the optimal choice of the critical 
mass in diminishing the magnitude of lattice artifacts at small quark mass 
can be found in~\cite{SHI,PAP} and in the work of refs.~\cite{CAN,XLFNEW}. 
As for Mtm-LQCD with clover-improved quark action, the promising quenched tests 
presented some years ago in~\cite{DMetal} have been recently extended 
in~\cite{Lub05} down to pion masses of 300~MeV or lower, confirming the 
absence of large cutoff effects. 

The outline of this presentation is as follows. In Section~\ref{sec:SEOLC} we 
analyze the form of the Symanzik expansion of lattice correlators beyond O($a$)
and explain why and how `IR divergent'' cutoff effects arise in this context.  
In section~\ref{sec:KLL} we discuss two ways of killing all the leading 
``IR divergent'' cutoff effects and we describe the structure of the 
left-over ``IR divergent'' terms. Finally in Section~\ref{sec:ART} 
we collect some remarks on the peculiar structure of the lattice artifacts 
affecting lattice hadronic energies and in particular pion masses.
Conclusions can be found in Section~\ref{sec:CONC}. 

\section{Symanzik analysis of ``IR divergent'' cutoff artifacts}
\label{sec:SEOLC}  

The study of cutoff artifacts affecting lattice correlators 
in Mtm-LQCD can be elegantly made in the language of the Symanzik 
expansion. A full analysis of cutoff effects beyond O($a$) is of 
course extremely complicated. Fortunately it is not necessary, 
if we limit the discussion to the terms that are enhanced as the 
quark mass $m_q$ is decreased.

$\bullet$ {\it The Symanzik LEEA of Mtm-LQCD} - The 
expression of the fermionic action of  Mtm-LQCD in the 
physical quark basis is given in~\cite{FR1,FR2}. The 
low energy effective action (LEEA), $S_{\rm Sym}$, of 
the theory can be conveniently written in the form
\beq
S_{\rm Sym}=\int\!d^4y\,\Big{[}{\cal L}_4(y)+
\sum_{k=0}^{\infty}a^{2k+1}\ell_{4+2k+1}(y)
+\sum_{k=1}^{\infty}a^{2k}\ell_{4+2k}(y)\Big{]}\, ,
\label{SLEEA}\eeq 
where ${\cal L}_4=\frac{1}{2g_0^2}{\rm tr}(F\!\cdot\! F)+ 
\bar\psi(\gamma \!\cdot\! D + m_q)\,\psi$ is the target 
continuum QCD Lagrangian. Based on the symmetries
of Mtm-LQCD a number of interesting properties 
enjoyed by $S_{\rm Sym}$ can be proved which are summarized below.

1. Lagrangian densities of even dimension, $\ell_{2k}$, in 
eq.~(\ref{SLEEA}) are parity-even, while terms of odd dimension, 
$\ell_{2k+1}$, are parity-odd and twisted in iso-spin space. 
Thus the latter have the quantum numbers of the neutral pion. 

2. The term of order $a$ in eq.~(\ref{DEFEO}), $\ell_5$, 
is given by the linear combination
\begin{eqnarray}
\hspace{-1.5cm}&&\ell_5=\delta_{5,SW}\,\ell_{5,SW}+\delta_{5,m^2}\,
\ell_{5,m^2}+\delta_{5,e}\,\ell_{5,e}\, ,\label{L5}\\
\hspace{-1.5cm}&&\ell_{5,SW}=
\frac{i}{4}\bar\psi[\sigma\cdot F]i\gamma_5\tau_3\psi \, ,
\!\!\quad\ell_{5,m^2} = m_q^2 \bar\psi i\gamma_5\tau_3\psi \, ,\!\!\quad
\ell_{5,e} = \Lambda_{\rm QCD}^2 \bar\psi i\gamma_5\tau_3\psi  \, ,
\label{L51}
\end{eqnarray}
where the coefficients $\delta_{5,SW}$, $\delta_{5,m^2}$ and $\delta_{5,e}$ 
are dimensionless quantities, odd in $r$. The operator $\ell_{5,e}$ arises 
from the need to describe order $a$ uncertainties entering any 
non-perturbative determination of the critical mass and goes together 
with $\ell_{5,SW}$. Both $\ell_{5,SW}$ and $\ell_{5,e}$ could be made to 
disappear from~(\ref{SLEEA}) by introducing in the Mtm-LQCD action the 
SW (clover)-term~\cite{SW} with the appropriate non-perturbatively 
determined $c_{SW}$ coefficient~\cite{LU} and at the same time setting 
the critical mass to its correspondingly O($a$) improved value. 

3. Higher order ambiguities ($k\geq 1$) in the critical mass, which 
will all contribute to ${\cal L}_{\rm odd}$, are described by terms 
proportional to odd powers of $a$ of the kind 
\beq a^{2k+1}\,\delta_{4+2k+1,e}\,\ell_{4+2k+1,e}=a^{2k+1}\,\delta_{4+2k+1,e}\,
(\Lambda_{\rm QCD})^{2k+2}\,\bar\psi i\gamma_5\tau_3\psi\, .\label{HK}\eeq

$\bullet$ {\it Describing Mtm-LQCD correlators beyond O($a$)} - 
We are interested in the Symanzik description of the lattice artifacts 
affecting connected expectation values of $n$-point, multi-local, 
multiplicative renormalizable (m.r.) and gauge-invariant operators
$O(x_1,x_2,\ldots,x_n)=\prod_{j=1}^{n} O_j(x_j)\equiv O(x)$,
$x_1\neq x_2 \neq \ldots \neq x_n$, which we take to have continuum 
vacuum quantum numbers, so as to yield a non trivially 
vanishing result as $a\to 0$. In order 
to ensure automatic O($a$) improvement~\cite{FR1} we shall assume that $O$ 
is parity invariant in which case its Symanzik expansion  
will contain only even powers of $a$. Schematically we write 
\beqn
\hspace{-.3cm}&&\<O({x})\>\Big{|}_{m_q}^{L}\!=\!
\<[O({x})+\Delta_{\rm odd}O(x)+\Delta_{\rm even}O(x)]
e^{-\int\!d^4 y[{\cal L}_{\rm odd}(y)+{\cal L}_{\rm even}(y)]}\>
\Big{|}^{\rm cont}_{m_q}\, , \label{SEOP}\\
&&{\cal L}_{\rm odd}=\sum_{k=0}^{\infty}a^{2k+1}\ell_{4+2k+1}\, ,\qquad
{\cal L}_{\rm even}=\sum_{k=1}^{\infty}a^{2k}\ell_{4+2k}\, .
\label{DEFEO}
\eeqn
The operators $\Delta_{\rm odd}O$ ($\Delta_{\rm even}O$)
have an expansion in odd (even) powers of $a$. They can be viewed as the 
$n$-point operators necessary for the on-shell improvement of 
$O$~\cite{HMPRS,LU}.

$\bullet$ {\it Pion poles and ``IR divergent'' cutoff effects} - 
Although a complete analysis of all the ``IR divergent'' cutoff effects 
is very complicated, the structure of the leading ones ($h=2k$ in 
eq.~(\ref{ORD})) is rather simple, as they only come from continuum 
correlators where $2k$ factors $\int d^4y\,{\cal L}_{\rm odd}(y)$ 
are inserted. More precisely the leading ``IR divergent'' cutoff 
effects are identified on the basis of the following result~\cite{FR4}. 

In the Symanzik expansion of $\<O(x)\>|^L_{m_q}$ at order $a^{2k}$ 
($k\geq 1$) there appear terms with a $2k$-fold pion pole and residues 
proportional to $|\<\Omega|{\cal L}_{\rm odd}|\pi^0({\bf 0})\>|^{2k}$, 
where $\<\Omega|$ and $|\pi^0({\bf 0})\>$ denote the vacuum 
and the one-$\pi^0$ state at zero three-momentum, respectively. 
Putting different factors together, each one of these terms can be seen 
to be schematically of the form (recall ${\cal L}_{\rm odd}={\rm O}(a)$) 
\beqn
\hspace{-1.cm}
\Big{[}\Big(\frac{1}{m_\pi^2}\Big)^{2k}
(\xi_{\pi})^{2k}{\cal M}[O;\{\pi^0({\bf 0})\}_{2k}]
\Big{]}_{m_q}^{\rm cont}\, ,\qquad
\xi_{\pi}=\Big{|}\<\Omega|{\cal L}_{\rm odd}|
\pi^0({\bf 0})\>\Big{|}_{m_q}^{\rm cont}\, ,\label{METREO}
\eeqn
where we have generically denoted by ${\cal M}[O;\{\pi^0({\bf 0})\}_{2k}]$ 
the $2k$-particle matrix elements of $O$, with each of the $2k$ particles 
being a zero three-momentum neutral pion. 

Less ``IR divergent'' cutoff effects (those with $h$ strictly 
smaller than $2k$ in eq.~(\ref{ORD})) come either from terms 
with some extra $\int d^4y{\cal L}_{\rm even}(y)$ insertions or from
contributions of more complicated intermediate states other than 
straight zero three-momentum pions or from both. In the first case 
one gets extra $a$ powers (not all ``compensated'' by corresponding 
pion poles), while in the second one loses some $1/m_\pi^2$ factor. 

It is important to remark that the appearance of pion poles like the 
ones in eq.~(\ref{METREO}) in no way means that the lattice correlators 
diverge as $m_q\to 0$, but only that the Symanzik expansion we have 
employed appears to have a finite radius of convergence (on this point 
see the remarks of ref.~\cite{SH05}).

\section{Reducing ``IR divergent'' cutoff artifacts}
\label{sec:KLL}

Recalling that ${\cal{L}}_{\rm odd}=a\,\ell_5+{\rm O}(a^3)$, the previous
analysis shows that at leading order 
in $a$ the residue of the most severe multiple pion poles is 
proportional to $|\<\Omega|\ell_{5}|\pi^0({\bf 0})\>|^{2k}$. It is an 
immediate conclusion then that the leading ``IR divergent'' cutoff 
effects can all be eliminated from lattice data if we can either reduce $\ell_{5}$ 
to only ${\ell}_{5,m^2}$ in~(\ref{L5}) or set $\xi_\pi$ to zero. 

$\bullet$ {\it Improving the Mtm-LQCD action by the SW-term} - 
The obvious, field-theoretical way to eliminate $\ell_{5}$ from 
the LEEA of Mtm-LQCD consists in making use of the O($a$) improved 
action~\cite{SW,LU,HMPRS}.
In this case lattice correlation functions will admit a Symanzik
description in terms of a LEEA where the operators ${\ell}_{5,SW}$ and
${\ell}_{5,e}$ are absent, and ${\ell}_5$ is simply given by 
${\ell}_{5,m^2}$. The left-over contributions 
arising from the insertions of ${\ell}_{5,m^2}$ in $\<O\>|_{m_q}^{\rm cont}$ 
yield terms that are at most of order 
$(am_q^2/m_\pi^2)^{2k}\simeq (a m_q)^{2k}$, hence negligible in the chiral 
limit. It is instead the next odd operator in the 
Symanzik expansion, $a^3\ell_7$, which comes into play. 

A detailed combinatoric analysis based on the structure of the 
non-leading ``IR divergent'' cutoff effects~\cite{FR4}
reveals that the worst lattice artifacts left behind in correlators 
after the ``clover cure'' are of the kind $a^2(a^2/m_\pi^2)^{k-1}$, $k\geq 1$. 

$\bullet$ {\it Optimal choice of the critical mass} - 
The alternative strategy to kill the leading ``IR divergent'' cutoff 
effects consists in leaving the Mtm-LQCD action unimproved,  
but fixing the critical mass through the condition
\beq 
\lim_{m_q \to 0^{+}} \xi_\pi(m_q)=\lim_{m_q \to 0^{+}} \;
\Big{|}\<\Omega|{\cal{L}}_{\rm odd}|\pi^0({\bf 0})\>
\Big{|}^{\rm cont}_{m_q}\; = \; 0 \, .
\label{EFFCOND}
\eeq 
The meaning of~(\ref{EFFCOND}) is simple. It amounts 
to fix, for $k\geq 0$, the order $a^{2k+1}$ contribution in the 
counter-term, $M_{\rm cr}\bar\psi^L i\gamma_5\tau_3\psi^L$, 
so that its vacuum to one-$\pi^0({\bf 0})$ matrix element  
compensates, in the limit $m_q\to 0$, the similar matrix element 
of the sum of all the other operators making up $\ell_{4+2k+1}$.

A concrete procedure designed to implement condition~(\ref{EFFCOND}) 
in actual simulations was discussed in ref.~\cite{FR4}. It consists in 
determining the critical mass by requiring the lattice correlator  
$a^3\sum_{\bf x}\;\<V_0^2(x)P^1(0)\>|^L_{m_q}$ ($x_0 \neq 0$)
to vanish in the chiral limit, where 
$V_0^2=\bar\psi\gamma_0\frac{\tau_2}{2}\psi$ is the vector current 
with iso-spin index 2 and $P^1=\bar\psi \gamma_5 \frac{\tau_1}{2}\psi$ 
the pseudo-scalar density with iso-spin index 1. 
In the continuum this correlator is zero by parity for any value of $m_q$.
On the lattice the breaking of parity (and iso-spin) due to the twisting 
of the Wilson term makes it non-vanishing by pure discretization artifacts, 
which have the form of a power series expansion in $\xi_\pi/m_\pi^2$. 
 
The important conclusion of the analysis presented in~\cite{FR4} 
is that it is not necessary (nor possible) to really go 
to $m_q\to 0$. It is enough to have the critical mass determined by the
vanishing of the above correlator 
at the current simulation quark mass, provided we stay in the region  
$m_q>a^2$. In these conditions we will have 
$\xi_\pi(m_q)={\rm O}(am_\pi^2)$ with all the leading ``IR divergent'' 
cutoff effects reduced to finite O($a^{2k}$) terms. As for the 
subleading ones, a non-trivial diagrammatic analysis shows that the worst 
of them, left behind after the ``optimal critical mass cure'', 
are reduced to only $a^2(a^2/m_\pi^2)^{k-1}$, $k\geq 1$, effects, just like 
in the case where the clover term is employed.

\section{Artifacts on hadronic energies and pion masses}
\label{sec:ART}

In the language of the Symanzik expansion discretization artifacts on 
hadronic energies are described by a set of diagrams where at least one
among the inserted $\int {\cal L}_{\rm odd}$ factors gets necessarily absorbed
in a multi-particle matrix element, with the consequence that it is
not available for producing a pion pole. 
As a consequence, at fixed order in $a$, 
the most ``IR divergent'' lattice corrections to continuum hadronic energies 
contain one overall factor $1/m_\pi^2$ less than the leading ``IR divergent'' 
cutoff effects generically affecting correlators.
For instance, to order $a^2$ the difference between lattice
and continuum energy of the hadron $\alpha_n$ reads~\cite{FR4}
\beqn
\hspace{-0.5cm}
\Delta E_{\alpha_n}({\bf q})\Big{|}_{a^2} \; \propto \; \left[
\frac{a^2}{m_\pi^2}
{\rm Re} \left(\frac{\<\Omega| \ell_5 | \pi^0({\bf 0}) \>
\<\pi^0({\bf 0})\alpha_n ({\bf q}) | \ell_5 |\alpha_n ({\bf q}) \>}
{2 E_{\alpha_n}({\bf q}) }\right)+ {\rm O}(a^2)
\right]_{m_q}^{\rm cont}\, ,
\label{DE2LEAD}
\eeqn
where ${\rm O}(a^2)$ denotes ``IR finite'' corrections. 
It should be noted that this ``IR divergent'' lattice artifact
is reduced to an ``IR finite'' correction
after anyone of the two ``cures'' described in Sect.~\ref{sec:KLL}. 
                                                                                        
Specializing the formula~(\ref{DE2LEAD}) to the case of pions, one
obtains the interesting result that the difference between charged
and neutral pion (square) masses is a finite O($a^2$) quantity even
if the critical mass has not been set to its optimal value or the
clover term has not been introduced. The reason is that the
leading ``IR divergent'' contributions shown in~(\ref{DE2LEAD})
are equal for all pions (as one can prove by standard soft pion
theorems~\cite{SPT}), hence cancel in the (square) mass difference.
This conclusion is in agreement with detailed results from chiral 
perturbation theory (see refs.~\cite{SCO} and~\cite{SHWUNEW}),
as well as with the first numerical estimates of the pion
mass splitting in Mtm-LQCD~\cite{LIV05}.

\section{Conclusions}
\label{sec:CONC}

When analyzed in terms of the Symanzik expansion, lattice correlators 
in Mtm-LQCD show ``IR divergent'' cutoff effects 
which tend to become large as the quark mass gets small. Extending 
the works of refs.~\cite{AB,SHWUNEW}, we have shown  
that, not only if the critical mass is chosen in some ``optimal'' way 
but also if the action is clover improved, such  
lattice artifacts are strongly reduced to terms that are at worst of the type 
$a^{2}(a^2/m_\pi^2)^{k-1}$, $k\geq 1$. 
The latter result implies that the continuum extrapolation of lattice data is smooth 
at least down to values of the quark mass satisfying the order of 
magnitude inequality $m_q >a^2\Lambda^3_{\rm QCD}$.

\vskip .2cm
\noindent{\bf Acknowledgments - } 
We thank the LOC for the exciting atmosphere of the 
Conference. G.C.R. gratefully acknowledges financial support from 
Humboldt Foundation and NIC (DESY - Zeuthen).


\begin{thebibliography}{99}

\bibitem{TM} 
R. Frezzotti, P.A. Grassi, S. Sint and P. Weisz, JHEP {\bf 0108} (2001) 058;\\
R. Frezzotti, S. Sint and P. Weisz [ALPHA Collaboration], JHEP {\bf 0107}
(2001) 048;\\
M. Della Morte, R. Frezzotti, J. Heitger and S. Sint [ALPHA Collaboration],
JHEP {\bf 0110} (2001) 041.

\bibitem{FR1}
R. Frezzotti and G.C. Rossi, JHEP {\bf 0408} (2004) 007 and 
JHEP {\bf 0410} (2004) 070;\\
R. Frezzotti, Nucl. Phys. {\bf B} (Proc. Suppl.) {\bf 140} (2005) 134.

\bibitem{FR2}
R. Frezzotti and G.C. Rossi, Nucl. Phys. {\bf B} (Proc. Suppl.)
{\bf 128} (2004) 193.

\bibitem{FR4}
R. Frezzotti, G. Martinelli, M. Papinutto and G.C. Rossi, 
hep-lat/0503034.  

\bibitem{SHWUNEW}
S.R. Sharpe and J.M.S. Wu, Phys. Rev. {\bf D71} (2005) 074501.

\bibitem{AB}
S. Aoki and O. B\"ar, Phys. Rev. {\bf D70} (2004) 116011. 

\bibitem{AB05} S. Aoki and O. B\"ar, these proceedings, hep-lat/0509002.

\bibitem{SH05} S.R. Sharpe,
hep-lat/0509009.

\bibitem{PS} S. Aoki, Phys. Rev. {\bf D30} (1984) 2653
and Phys. Rev. Lett. {\bf 57} (1986) 3136;\\
%
S.R. Sharpe and R. Singleton, Jr., Phys. Rev. {\bf D58} (1998) 074501;\\
%
G. M\"unster, JHEP {\bf 0409} (2004) 035;\\
S.R. Sharpe and J.M.S. Wu, Phys. Rev. {\bf D70} (2004) 094029.

\bibitem{SCO}
L. Scorzato, Eur. Phys. J. {\bf C37} (2004) 445.

\bibitem{SHI}
A. Shindler, these Proceedings.

\bibitem{PAP}
M. Papinutto, these Proceedings.
                                                                                       
\bibitem{CAN}
A.M. Abdel-Rehim, R. Lewis and R.M. Woloshyn, Phys. Rev. {\bf D71}
(2005) 094505.
                                                                                       
\bibitem{XLFNEW}
K. Jansen {\it et al.},
[$\chi$LF Collaboration], Phys. Lett. {\bf B619} (2005) 184
and hep-lat/0507010.

\bibitem{DMetal} 
M. Della Morte, R. Frezzotti and J. Heitger [ALPHA Collaboration],
  Nucl. Phys. {\bf B} (Proc. Suppl.) {\bf 106} (2002) 260 and hep-lat/0111048.  

\bibitem{Lub05} 
V. Lubicz, these Proceedings.

\bibitem{SW}
B. Sheikholeslami and R. Wohlert, Nucl. Phys. {\bf B259} (1985) 572.

\bibitem{LU}
M. L\"{u}scher, S. Sint, R. Sommer and P. Weisz, Nucl. Phys. {\bf B478}
(1996) 365;\\
M. L\"{u}scher, S. Sint, R. Sommer, P. Weisz and U. Wolff, Nucl. Phys.
{\bf  B491} (1997) 323.

\bibitem{HMPRS}
G. Heatlie, G. Martinelli, C. Pittori, G.C. Rossi and 
C.T. Sachrajda, Nucl. Phys. {\bf B352} (1991) 266.

\bibitem{SPT}
S. Weinberg, Phys. Rev. {\bf D7} (1973) 1068.

\bibitem{LIV05}
K. Jansen {\it et al.} [$\chi$LF Collaboration],
hep-lat/0507032;\\
F. Farchioni {\it et al.},
hep-lat/0509036.

\end{thebibliography}
\end{document}